\documentclass[aps,prc,twocolumn,nofootinbib]{revtex4-2}

\usepackage{mathptmx}
\usepackage{graphicx}
\usepackage[tbtags]{amsmath}
\usepackage{bm,amssymb}
\newcommand{\bder}{{\bm{\nabla}}}
\newcommand{\br}{{\bm{r}}}
\newcommand{\bp}{{\bm{p}}}

\newcommand{\Tr}{\mathop{\rm Tr}}

\newcommand{\nuc}[2]{{\mbox{$^{#2}${#1}}}}
\def\<{\langle}
\def\>{\rangle}
\renewcommand{\emph}[1]{\textit{#1}}

\begin{document}
\title{Semiclassical origin of nuclear ground-state octupole
deformations}
\author{Ken-ichiro Arita\email{arita@nitech.ac.jp}}
\affiliation{Department of Physics, Nagoya Institute of Technology,
Nagoya 466-8555, Japan}
\date{October 18, 2023}

\begin{abstract}
\begin{description}
\item[Background]
Ground-state octupole deformations are suggested in nuclei located
in the \textit{north-east} neighbor of the doubly magic nuclei
on the nuclear
chart $(N,Z)$, such as those in Ba and Ra-Th regions.
This systematics has been attributed to the parity mixing of the
approximately degenerate $\varDelta l=3$ pair of single-particle
levels near the Fermi surface.
\item[Purpose]
Nuclear deformations are governed in most cases by the gross shell
structures of the single-particle spectra.
I will consider the systematics in octupole deformation
from the view point of the gross shell structure, and investigate
the mechanism of its manifestation using the semiclassical
periodic-orbit theory (POT), which describes the quantum shell
effect by means of the periodic orbits (POs) in the corresponding
classical system.
\item[Methods]
To focus on the role of deformation, simplified infinite-well (cavity)
and radial power-law potential models are employed taking account of
quadrupole and octupole shape degrees of freedom.  Nuclear
ground-state deformations are investigated over the nuclear chart, and the
properties of the deformed shell structures are analyzed by means of
the semiclassical POT.
\item[Results and conclusions]
The systematics in nuclear ground-state octupole deformations are
reproduced in simplified mean-field potential models either with or without
parity mixing between $\varDelta l=3$ pair of levels.  The strong
octupole deformed shell effect at above the spherical shell closures
are explained simply and clearly using the semiclassical
POT.  They are associated with the local restoration
of dynamical symmetry, which enhance the contribution of classical
POs to the gross shell effect.
\end{description}

\end{abstract}
\maketitle

\section{Introduction}

Breaking of reflection symmetry is one of the fundamental problems in
nuclear structure physics \cite{Butler96}.  In medium-mass to heavy
nuclei, ground-state octupole deformations are observed only for a few
isotopes such as those around neutron-rich Ba region and Ra-Th region.
Possible static octupole shapes for even-even nuclei have been
systematically investigated over the nuclear chart by means of various
theoretical approaches
\cite{Moller08,Robledo11,Agbem16,Ebata17,Cao20}.  All those studies
have obtained the results which are basically consistent with the
experiments.

As well as the ground-state deformation, the significance of the
octupole shape degree of freedom in nuclear fission has been also
suggested \cite{Scamps18,Scamps19}.  The reason why the
fission-fragment mass distribution of actinide nuclei is centered at
$A\sim 140$, which are slightly larger than that of the
doubly magic \nuc{Sn}{132}, can be understood by considering the
shell effect of the
pare-shaped prefragment.  The octupole shape degree of
freedom should also play role
in the process of superasymmetric fission, referred to as cluster
radioactivity, where the shell effect of doubly magic
\nuc{Pb}{208} is concerned \cite{Warda12,Warda18,Mathe19}.

It has been considered that the nuclear octupole deformations are
attributed to the octupole correlation
between the approximately degenerate $\varDelta
l=3$ pair of single-particle levels near the Fermi surface.
Such pairs of degenerate levels arise just above each
spherical shell gaps due to the intruder levels from the higher oscillator
shells in the realistic nuclear mean-field potential with sharp
surface \cite{Butler96}.
For instance, proton $(h_{11/2},d_{5/2})$
levels above $Z=50$ gap and neutron $(i_{13/2},f_{7/2})$ levels above
$N=82$ gap are thought to be relevant for the octupole softness
in Ba region.

In addition to the above $\varDelta l=3$ mixing, I have pointed out
the significance of the gross shell effect for the octupole
deformation \cite{Arita23}.  In that work, infinite-well potential
(cavity) model was employed where the surface shape is parametrized by
merging a sphere and a paraboloid.  The semiclassical periodic-orbit
theory (POT) \cite{BaBlo3,Gutz71} is successfully utilized to elucidate
the origin of
remarkable shell structure for octupole deformed nuclei.
In POT, quantum shell effect is described by means of the
periodic orbits (POs) in the corresponding classical system.
The advantage of such cavity model is that the contribution of the
classical POs to the shell
energies can be obtained by directly evaluating the semiclassical
trace formula, which represents the quantum level density (density
of energy eigenvalues) as the sum
over contributions of the classical POs.  For the
system with a few particles added to the spherical closed-shell
configurations, octupole shape is advantageous in gaining large shell
energy, and its reason can be clearly explained using the contribution
of degenerate family of classical POs
to the semiclassical density of states.

In this study, I extend the above model a little to consider the
shapes with arbitrary combinations of axially symmetric quadrupole and
octupole deformations.  With this extremely simplified mean-field
model,
I would like to focus only on the effect of deformation.  
The central aim of this work is to investigate the role of the gross
shell effect to the systematic appearance of octupole deformations
above the spherical closed-shell configurations, and clarify their
origin by the semiclassical POT.
In my previous studies with my collaborators on
the cavity and oscillator-type potential models,
it has been shown that
the bifurcations of equatorial orbits at certain combinations of
axially-symmetric quadrupole and octupole deformations provide
remarkable shell effects \cite{Arita95,Sugita97}.  The PO bifurcations
is associated with the local restoration of symmetry, with which the
family of classical POs acquire extra local degeneracies.
Since the oscillator type potential has no $\Delta l=3$ pairs of
levels
at the spherical shape, it may also give us information about the
relative importance of $\Delta l=3$ mixing in octupole deformation.

This paper is organized as follows.
In Sec.~\ref{sec:theory}, a brief review on the semiclassical theory
of single-particle shell structure is given.  In Sec.~\ref{sec:param},
the mean-field potential models employed in this work are defined.  In
addition to the traditional prescription to expand the surface shape
by spherical harmonic functions, a specific way of parametrization is
proposed by merging a spheroid and a paraboloid.  In the cavity model
with the latter parametrization, classical POs form
continuous families with higher degeneracy and a stronger deformed
shell effect is expected.  Then, systematic calculations of
ground-state deformations over the nuclear chart is performed in
Sec.~\ref{sec:calc} for both parametrizations above.  The condition
for nuclei to gain shell energy by octupole deformation is
considered using the relation between the gross shell structure and
classical POs, and the mechanism for the systematic
appearance of octupole deformation at above the spherical
closed-shell configurations is explained.
It will be also shown that the above systematics is reproduced in the
oscillator type potential model, namely, without the help of $\Delta
l=3$ mixing.  Section~\ref{sec:summary} is devoted to the summary and
concluding remarks.

\section{Theoretical framework}
\label{sec:theory}

\subsection{Periodic orbit theory}
\label{sec:pot}

Semiclassical periodic-orbit theory (POT) is the powerful tool to
analyze the gross shell structures \cite{Gutz71,BaBlo3,BBBook}, and I
have taken full advantage of it in investigating the microscopic
origin of nuclear deformations and shape
stabilities \cite{Arita12,Arita16}.  Here, let us briefly review some
of the key issues related to the POT.

When one solves the quantum single-particle energy eigenvalue problem,
the distribution of the energy levels generally show a regular
oscillating pattern.  However, the origin for this structure
generally cannot be
explained by purely quantum-mechanical concepts alone.
Using the semiclassical approximation to the path-integral
representation of the Green's function, contribution of classical POs
are extracted, and the level density
\begin{equation}
g(e)=\sum_{i}\delta(e-e_i)
\end{equation}
is expressed as the sum over the contribution of classical POs
\begin{gather}
g(e)=\bar{g}(e)+\delta g(e), \nonumber \\
\delta g(e)\simeq \sum_{\rm PO}A_{\rm PO}(e)\cos\left(\tfrac{1}{\hslash}
 S_{\rm PO}(e)-\tfrac{\pi}{2}\mu_{\rm PO}\right), \label{eq:trace_g}
\end{gather}
which is known as the trace formula \cite{Gutz71,BBBook}.
The average part $\bar{g}(e)$ is given by the (extended) Thomas-Fermi
approximation.  In the oscillating part $\delta g$, $S_{\rm PO}=\oint_{\rm
PO}\bp\cdot d\br$ represents the action integral along the PO,
$\mu_{\rm PO}$ is the Maslov index related to the geometrical
character of the orbit, and the amplitude $A_{\rm PO}$ is determined
by the degeneracy, period and stability of the PO.  Since the action
integral is generally a monotonically increasing function of energy
$e$, each contribution of PO in Eq.~(\ref{eq:trace_g}) gives a
regularly oscillating function of $e$.  The orbit with shorter period
$T_{\rm PO}=dS_{\rm PO}/de$ gives the gross structure of the level
density and the longer orbits contribute to the finer structures.  In
consideration of gross shell structure, one has only to take the
contributions of a few shortest POs.  Under continuous symmetries, the
orbits will form a continuous family.  This is called a degeneracy of
the classical POs.  The orbit with higher
degeneracies make more significant contribution to the level density
in the $\hslash$ expansion.  The number of continuous parameters
$K_{\rm PO}$ for the PO family is called the degeneracy parameter, and
the amplitude factor $A_{\rm PO}$ is proportional to $\hslash^{-K_{\rm
PO}/2}$.

Using Eq.~(\ref{eq:trace_g}), one obtains the trace formula for
shell energy as \cite{Strut76,BBBook}
\begin{align}
&\delta E(N)=\int^{e_F}(e-e_F)\delta g(e)de \nonumber \\
&~ \simeq\sum_{\rm PO}\frac{\hslash^2}{T_{\rm PO}^2}
A_{\rm PO}(e_F)\cos\left(\tfrac{1}{\hslash}S_{\rm PO}(e_F)
-\tfrac{\pi}{2}\mu_{\rm PO}\right), \label{eq:trace_sce}
\end{align}
where $e_F$ is the Fermi energy satisfying
\begin{equation}
N=\int^{e_F}g(e)de.
\end{equation}
Due to the additional factor $T_{\rm PO}^{-2}$ in the amplitudes
of the PO contributions, longer orbits become less important,
and accordingly, one has only to consider the POs that are short
and preferably of higher degeneracies.

Another important aspect of the PO contribution is related
to the stability of the orbits.  The amplitude factor $A_{\rm PO}$
is proportional to the stability factor as follows:
\begin{equation}
A_{\rm PO}\propto \frac{1}{\sqrt{|\det(I-\tilde{M}_{\rm PO})|}},
\label{eq:stability}
\end{equation}
where $\tilde{M}$ represents the symmetry-reduced
monodromy matrix which describes the linear stability of the orbit.
In calculating the monodromy matrix, one sets a $2f-2$ dimensional
``surface of section'' $\Gamma$ in the classical $2f-1$ dimensional
phase space
with energy constraint $H(\br,\bp)=E$, where $f$ is the number of
degrees of freedom.
Then, consider a trajectory starting off at $Z_0$ on the
surface $\Gamma$.  Since the energy surface
$H(\br,\bp)=E$ is compact, the trajectory
will intersect the surface $\Gamma$ again at $Z_1$ in the same
direction.  The successive plots of the intersection
points $Z_1, Z_2,\cdots$ is called a Poincar\'{e} surface of section
(PSS) plot.  The map $\mathcal{M}$ from $Z_k$ to $Z_{k+1}$
$[Z_{k+1}=\mathcal{M}(Z_k)]$ defined by the Hamiltonian dynamics is
called the Poincar\'{e} map.
The PO is nothing but the fixed point $Z^*$ of the Poincar\'{e}
map $\mathcal{M}$ (or its power $\mathcal{M}^n$ in general),
satisfying $Z^*=\mathcal{M}^n(Z^*)$.
\begin{figure}
\centering
\includegraphics[width=.4\linewidth]{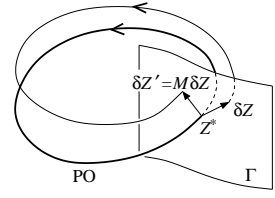} 
\caption{\label{fig:monodromy}
Calculation of monodromy matrix for PO on the surface $\Gamma$.}
\end{figure}
As shown in Fig.~\ref{fig:monodromy},
let us consider a trajectory near the PO, staring off at
$Z_0=Z^*+\delta Z$ on $\Gamma$.  It is generally nonperiodic and will
arrive at $Z_n=Z^*+\delta Z'$ on $\Gamma$, and $\delta Z'$ can be
written as
\begin{equation}
\delta Z'=M_{\rm PO}\delta Z + O(\delta Z^2).
\end{equation}
The $(2f-2)\times(2f-2)$ matrix $M_{\rm PO}$, representing
a linearized Poincar\'{e} map, defines the monodromy
matrix.  $M_{\rm PO}$ depends on the choice of $\Gamma$, but its
eigenvalues are irrelevant to $\Gamma$ and therefore the stability
factor (\ref{eq:stability}) does not depend on $\Gamma$.

For system with continuous
symmetries, $M_{\rm PO}$ has trivial unit eigenvalues, and the
symmetry-reduced matrix $\tilde{M}_{\rm PO}$ is obtained by splitting
off the degrees of freedom relevant to those symmetries.  It can
happen that one of the eigenvalues of $\tilde{M}_{\rm PO}$
becomes equal to 1 with varying parameter of the Hamiltonian.  This
corresponds to the bifurcation point of PO, where dynamical symmetry
is locally restored around the PO and the orbit forms a local
continuous family.  The
orbits belonging to such a family make coherent contribution to the
level density and brings about a significant enhancement of the
amplitude factor $A_{\rm PO}$.
Since the standard stationary
phase approximation (SPA) to derive the trace formula is broken down at the
bifurcation point, the stability factor in
Eq.~(\ref{eq:stability}) suffers divergence there.
This shortcoming can
be remedied by an appropriate treatment of the higher-order expansions
about the PO, e.g., by the uniform
approximations \cite{Schom97,Sieber96,Ozorio87}.
Bifurcation of short PO is often responsible for the emergence of
significant shell effect at exotic shapes.  This feature plays an
essential role when I consider the quadrupole-octupole deformations
in the following part.

\subsection{The shell-deformation energy}

When one employs an effective mean-field model, shell
energy is extracted from the single-particle spectra by
decomposing the sum of single-particle energies into the smooth and
oscillating parts as
\begin{gather}
E_{\rm sp}(Z,N;q)=\sum_{i=1}^Ze_i(q)+\sum_{j=1}^Ne_j(q) \nonumber \\
=\tilde{E}_{\rm sp}(Z,N;q)+\delta E(Z,N;q).
\end{gather}
$e_i(q)$ represents the single-particle energy for deformation $q$.
In the microscopic-macroscopic model, the oscillating part of the
single-particle energy sum is
added to the semi-empirical liquid-drop model (LDM) energy as
\begin{equation}
E(Z,N;q)=E_{\rm LDM}(Z,N;q)+\delta E(Z,N;q)
\end{equation}
In the present work, the employed mean field is not a realistic one, and
the use of a realistic LDM is of no importance.
Assuming the single-particle Hamiltonian $h=t+u$ ($t$ and $u$ being
kinetic energy and mean-field potential, respectively) as what is deduced
from the many-body Hamiltonian with two-body interaction, the smooth
part of the total energy is expressed as
\begin{equation}
\tilde{E}=\sum_{i=1}^N(\<t_i\>+\tfrac12\<u_i\>).
\label{eq:e_smooth}
\end{equation}
The factor $\frac12$ in the second term above is to avoid the double
counting of the interaction.  When the radial power-law potential
$u\propto r^\alpha$ is employed as the mean-field potential, the
Virial theorem gives the relation
\begin{equation}
\<t\>=\frac12\<\br\cdot\bder u\>=\frac{\alpha}{2}\<u\>.
\end{equation}
Together with the relation
$\<h\>=\<t\>+\<u\>$, one obtains
\begin{gather*}
\<t\>=\frac{\alpha}{\alpha+2}\<h\>, \quad
\<u\>=\frac{2}{\alpha+2}\<h\>.
\end{gather*}
Inserting them into Eq.~(\ref{eq:e_smooth}), one has
\begin{equation}
\tilde{E}
=\frac{\alpha+1}{\alpha+2}\sum_{i=1}^N\<h_i\>
=\frac{\alpha+1}{\alpha+2}\tilde{E}_{\rm sp}.
\end{equation}
Consequently, the total energy can be expressed as
\begin{equation}
E(Z,N;q)=\frac{\alpha+1}{\alpha+2}\tilde{E}_{\rm sp}(Z,N;q)+\delta E(Z,N;q)
\label{eq:energy_rpl}
\end{equation}
In the cavity limit, $\alpha\to\infty$,
one simply has
\begin{equation}
E^{\rm(cavity)}(Z,N;q)=E_{\rm sp}(Z,N;q).
\label{eq:energy_cavity}
\end{equation}
Ground-state deformation $q^*$ is obtained by minimizing the total
energy $E$ with respect to $q=\{q_2,q_3\}$,
\begin{equation}
E_{\rm min}(Z,N)=E(Z,N;q^*)=\min_{q_2,q_3} E(Z,N;q).
\end{equation}

In the analysis of deformation, one has usually considered the deformation
energy $E_{\rm def}$ which is defined
with the energy at spherical shape as reference;
\begin{equation}
E_{\rm def}(Z,N;q)=E(Z,N;q)-E(Z,N;0).
\label{eq:energy_def}
\end{equation}
In this definition, the reference energy is a fluctuating function
of particle numbers $Z$ and $N$.
To investigate the nuclear energy from a more general point of
view, without special reference to quantum fluctuation at the
spherical shape, 
let us define the {\it shell-deformation energy}, $E_{\textrm{sh-def}}$,
by estimating the total energy with the smooth part of the
spherical energy as reference;
\begin{align}
&E_{\textrm{sh-def}}(N,Z;q)=E(Z,N;q)-\tilde{E}(Z,N;0)
 \nonumber \\
&\quad =\{\tilde{E}(Z,N;q)-\tilde{E}(Z,N;0)\}
 +\delta E(Z,N;q).
\label{eq:energy_sh_def}
\end{align}
As seen from the right-hand side, it consists of the shell energy and
the smooth (LDM) deformation energy.  In the following analysis, the
shell-deformation energy (\ref{eq:energy_sh_def}) shall be referred
rather than the traditional deformation energy (\ref{eq:energy_def}).

\section{Shape parametrization with octupole and quadrupole
deformations}
\label{sec:param}

\subsection{Stretched octupole parametrization}

Various ways of parametrizing the shape of the nuclear
surface have been proposed to discuss the nuclear deformation.
The most commonly used one might be the expansion by spherical
harmonic functions.  For axially symmetric shapes, the surface
shape is expressed as
\[
r=r_s(\theta)=R_0\biggl[1+\sum_{l\ge 2}\beta_l P_l(\cos\theta)\biggr],
\]
where $P_l(x)$ is the Legendr\'{e} polynomial.  The $P_1$ term is also
considered when one wants to exactly eliminate the center of mass.

\begin{figure}
\centering
\includegraphics[width=\linewidth]{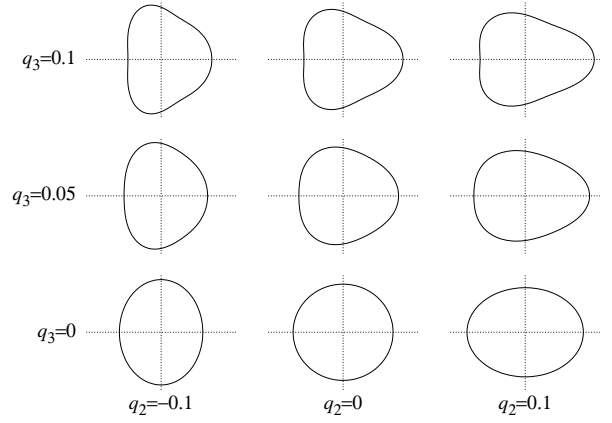} 
\caption{\label{fig:shapes_oct}
Quadrupole-octupole surface by the stretched octupole parametrization
$r=R_0f(\theta)$ [Eq.~(\ref{eq:shape_oct})] at several values of
quadrupole and octupole moments $(q_2,q_3)$.  Horizontal line is the
symmetry axis, and the origin is the center of mass.}
\end{figure}

In my recent studies, the above shape function is modified a little.
Instead of quadrupole deformation described by $P_2$ function,
I take the spheroidal shape
\[
r=r_s(\theta)=R_0\frac{\eta^{2/3}}{\sqrt{\cos^2\theta+\eta^2\sin^2\theta}},
\]
where $\eta=r_s(0)/r_s(\frac{\pi}{2})$ represents the axis ratio.
$\eta>1$ and $\eta<1$ correspond to
prolate and oblate deformations, respectively.  Deformations with
higher multipoles are considered by multiplying $r_s$ by an
exponential function $e^{\beta_lP_l(\cos\theta)}$.
They are first taken on the spherical surface, and then stretched (or
contracted) in the direction of the symmetry axis.  For the octupole
deformation, the surface profile function $f(\theta)=r_s(\theta)/R_0$
is expressed as
\begin{gather}
f(\theta)=c(\beta_3)\eta^{-1/3}\sqrt{1+(\eta^2-1)\cos^2\theta'}\,
e^{\beta_3P_3(\cos\theta')}, \nonumber \\
\tan\theta'=\eta\tan\theta.
\label{eq:shape_oct}
\end{gather}
$c_3(\beta_3)$ is given by
\begin{equation}
c(\beta_3)=\left[\frac12\int_{-1}^1 e^{3\beta_3P_3(t)}dt\right]^{-1/3}.
\end{equation}
so that the volume conservation condition is satisfied.
The center of mass condition is satisfied up to the first order of $\beta_3$
for a uniform rigid body with this surface.
By using the above exponential form, a natural surface shape can be
achieved up to rather large octupole deformations
(See Fig.~\ref{fig:shapes_oct}).

In displaying the potential-energy surface, I use the dimensionless
quadrupole and octupole moments as the shape parameters.
They are defined by
\begin{align}
q_2=\frac{1}{R_0^2}\<r^2P_2(\cos\theta)\>, \quad
q_3=\frac{1}{R_0^3}\<(r^3P_3(\cos\theta))'\>,
\label{eq:qmom}
\end{align}
where $(r^nP_n)'$ is the stretched multipole operator defined by the
stretched coordinate
\begin{equation}
(x',y',z')=(\eta^{1/3}x,~ \eta^{1/3}y,~ \eta^{-2/3}z).
\label{eq:scoord}
\end{equation}
By defining the octupole moment $q_3$ in terms of the stretched
coordinate as above, it becomes independent of $\eta$ and has
one-to-one correspondence with the parameter $\beta_3$.  The shapes of
the surface at several values of $(q_2,q_3)$ are displayed in
Fig.~\ref{fig:shapes_oct}.

\begin{figure}
\includegraphics[width=.8\linewidth]{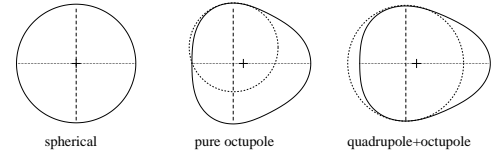} 
\caption{\label{fig:pict1}
Illustration of the local restoration of spherical symmetry in
quadrupole-octupole cavity potential model.  Horizontal line
represents the symmetry axis and the cross represents the center of
mass.  The vertical line represents the plane referred as the
equatorial plane.  The circle with dashed line represents the circle
of curvature on the meridian plane.  In the right panel, the shape of
a special combination of quadrupole and octupole deformation is shown
where the meridian curvature radius coincides with the equatorial
radius.}
\end{figure}

In the cavity potential with the surface $r=R_0f(\theta)$, one finds
regular polygon orbits on the plane perpendicular to the symmetry
axis, which I have called {``the equatorial plane''}, although it is
slightly translated from the $\theta=\frac{\pi}{2}$ plane for
$\beta_3\ne 0$.  Those orbits form degenerate one-parameter family
with respect to the rotation around the symmetry axis.  By varying the
deformation parameters, two main curvature radii at the equatorial
plane coincide with each other for certain combinations of
$(\eta,\beta_3)$ or $(q_2,q_3)$ (see right panel of
Fig.~\ref{fig:pict1}).
For such shapes, spherical symmetry is locally restored around the
equatorial plane, and the orbits on it acquire two extra
local degeneracies.
This condition is satisfied approximately along the $q_3=q_2$ line
on the shape parameter space.

The local symmetry restoration as discussed above provides substantial
enhancement of the shell effect and plays significant roles in
deformations of the systems \cite{Sugita97}.  A similar local symmetry
restoration is also found in smooth potential models, where the symmetries
are of the dynamical ones associated with the PO
bifurcations \cite{Arita95}.

\subsection{Spheroid + paraboloid parametrization}

\begin{figure}
\includegraphics[width=.6\linewidth]{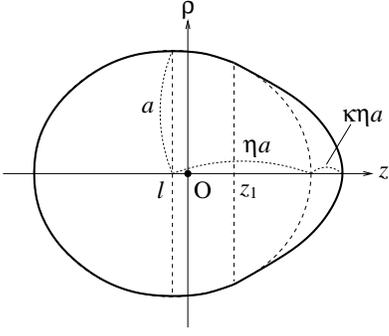} 
\caption{\label{fig:wqs}
Parametrization of quadrupole-octupole surface
by merging a spheroid and a paraboloid.}
\end{figure}

Next, I propose another way of shape parametrization.  It is known
that the spheroidal cavity model has nontrivial dynamical
symmetry \cite{Mag02}, and
the classical POs form continuous families with higher
degeneracy than the other axial shapes.  Since the orbit family of
higher degeneracy has stronger contribution to the quantum shell
effect, it would be advantageous for the octupole deformed surface
to include the spheroidal part in it.  For this reason, let us consider
axially symmetric octupole surface by merging a spheroid and a
paraboloid as shown in Fig.~\ref{fig:wqs}.
In the cylindrical coordinate $(\rho,z,\varphi)$, the
surface shape $\rho=\rho_s(z)$ is expressed as
\begin{gather}
\rho_s^2(z)=\left\{\begin{array}{l@{\quad}l}
 a^2-\left(\frac{z-l}{\eta}\right)^2 &
 (z_{\rm min}\leq z\leq z_1) \\
 2b(z_{\rm max}-z) & (z_1\leq z\leq z_{\rm max})
                   \end{array}\right., \label{eq:shape_wqs} \\
z_{\rm min}=l-\eta a, \quad
z_{\rm max}=l+(1+\kappa)\eta a. \nonumber
\end{gather}
A spheroidal surface $(z<z_1)$ and a paraboloidal surface $(z>z_1)$
are smoothly joined at $z=z_1$.
The shape of the entire surface is controlled by the two independent
shape parameters ($\eta$, $\kappa$).  $\eta$ is the axis ratio of
the spheroidal part, and $\kappa$ is the octupole parameter defined by
the relative width of the paraboloidal part raised from the spheroidal
surface.  For $\kappa=0$, there is no paraboloidal part and the entire
surface is the pure spheroid.  Especially, the entire surface is
spherical at $(\eta,\kappa)=(1,0)$.  The four other parameters $a$,
$b$, $l$ and $z_1$ entering in Eq.~(\ref{eq:shape_wqs}) are determined
by (i) the condition to merge the two surfaces smoothly at $z=z_1$,
(ii) the volume-conservation condition, and (iii) the center-of-mass
condition.  The center-of-mass condition does not affect the energy
eigenvalues, but is necessary in obtaining the correct quadrupole and
octupole moments.

\begin{figure}
\centering
\includegraphics[width=\linewidth]{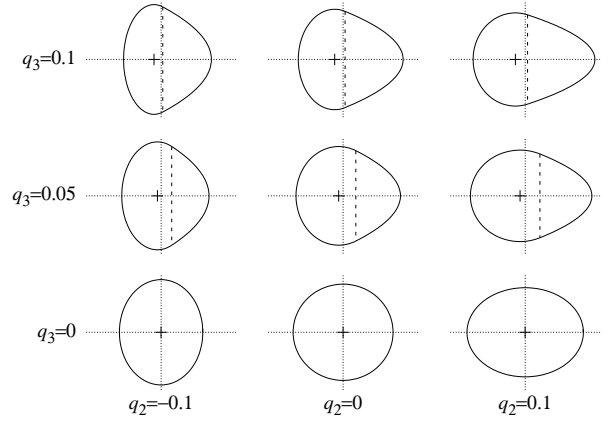} 
\caption{\label{fig:shapes_wqs}
Quadrupole-octupole surface by the spheroid-paraboloid parametrization
$\rho=\rho_s(z)$ [Eq.~(\ref{eq:shape_wqs})] at several values of
quadrupole and octupole moments $(q_2,q_3)$.  The cross represents
the center of the spheroid, and the broken line represents the
boundary of the spheroidal and paraboloidal surfaces, $z=z_1$.}
\end{figure}

In this shape parametrization, the deformed shell structures are
also investigated as functions of quadrupole and octupole moments ($q_2,
q_3$) defined by Eq.~(\ref{eq:qmom}).  In this case, the use of the
stretched coordinate in Eq.~(\ref{eq:qmom}) guarantees that the
octupole moment $q_3$ is independent on $\eta$ and has one-to-one
correspondence with the parameter $\kappa$.  The shape of the surface
at several values of $(q_2,q_3)$ are displayed in
Fig.~\ref{fig:shapes_wqs}.

In the cavity potential with the surface (\ref{eq:shape_wqs}), one
expects a strong shell effect when the spheroidal part of the surface
is spherical ($\eta=1$).  For such shape, the equatorial orbits on the
plane $z=l$ that are one-parameter families for $\eta\ne 1$ becomes
three-parameter families, which are not only local ones but exist over
finite ranges of the rotational angles.  This condition is satisfied
for smaller value of $q_2$ for a given $q_3$, compared with the last
parametrization.  In my recent analysis \cite{Arita23}, gross shell
structures similar to that for the spherical potential are found to
survive up to rather large octupole parameter $\kappa$ keeping
$\eta=1$.

\section{Systematics of ground-state octupole deformations
in cavity and oscillator potential models}
\label{sec:calc}

\subsection{Semiclassical mechanism of octupole deformation}

Based on the semiclassical trace formula, let us consider the condition
where the system acquires large shell energy gain by the octupole
deformation.  As discussed above, local symmetry restoration with a
special combination of quadrupole and octupole deformations brings
about a strong shell effect.
At those shapes, the POs with extra degeneracies are expected to
make similar contribution to the shell energy as in the
spherical potential.

In the cavity potential model, momentum $|\bp|=\hslash
k$ is constant and the action integral is given by the product of
momentum and the geometric length $L_{\rm PO}$.  The trace formula
(\ref{eq:trace_sce}) is then expressed as
\begin{equation}
\delta E(N)=\sum\frac{\hslash^2}{T_{\rm PO}^2}A_{\rm PO}
\cos(k_FL_{\rm PO}-\tfrac{\pi}{2}\mu_{\rm PO})
\label{eq:trace_cavity}
\end{equation}
where $k_F$ represents the Fermi wave number.

Because of the saturation property, the volume $V$ surrounded by the
potential surface is proportional to the particle number $N$.
According to the Weyl's formula \cite{BaBlo1}, the
level density in terms of the variable $e=k^2$ is given by
\begin{equation}
\rho(e)=\frac{Vk}{4\pi^2}
\end{equation}
with the volume
\begin{equation}
V=\frac{4\pi}{3}R_0^3=\frac{4\pi}{3}Nr_0^3 \quad
(R_0=r_0N^{1/3}).
\end{equation}
From the relation between Fermi wave number $k_F$ and particle number $N$,
one obtains
\begin{gather*}
N=\int_0^{k_F^2}\rho(e)de=\frac{Vk_F^3}{6\pi^2}
 =\frac{2N(k_Fr_0)^3}{9\pi}, \\
k_F=\left(\frac{9\pi}{2}\right)^{1/3}r_0^{-1}.
\end{gather*}
Thus, the value of the Fermi wave number $k_F$ is approximately
independent on the particle number $N$.
Consequently, the shell energy minimization condition associated with
the contribution of PO in Eq.~(\ref{eq:trace_cavity}),
\begin{equation}
\cos(k_FL_{\rm PO}-\pi\mu_{\rm PO}/2)=-1,
\end{equation}
is governed solely by the length of the orbit $L_{\rm PO}$
if the change of the Maslov index is ignored.
Concerning to a specific PO, the above condition
is that the radius $a$ of the equatorial plane to coincide with the radius
$R_0$ of the spherical magic nucleus.

Let $N_0$ a spherical magic number and $R_0$ the radius of the
surface for this magic nucleus.  For the octupole shape $q^*$ with
local symmetry ($q_3\approx q_2$ for stretched octupole, and $\eta=1$ for
spheroid + paraboloid parametrization), the radius $a$ of the equatorial
plane is shorter than the
radius of the sphere with the same volume.  In order for the radius $a$
to coincide $R_0$ which satisfies the energy minimization condition, the
particle number $N$ should be larger than $N_0$ as
\begin{gather}
a(N,q^*)=R_0=\left(\frac{N_0}{N}\right)^{1/3}a(N,0), \nonumber \\
N=N_0\left(\frac{a(N,0)}{a(N,q^*)}\right)^3
\label{eq:cond_emin}
\end{gather}
Thus, with increasing octupole deformation $q^*$, the energy minimization
condition is satisfied in the system with particle number
slightly larger than the spherical magic number.  This 
argument provides a simple and clear explanation for the octupole
deformation to be found
systematically at just above the spherical magic numbers.

\begin{figure}
\includegraphics[width=.8\linewidth]{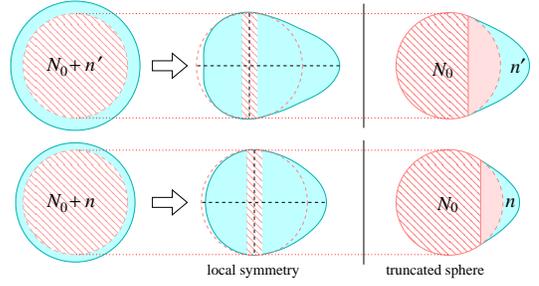} 
\caption{\label{fig:pict2}
Conceptual illustration of the mechanism of octupole deformation for
a systems with $N_0+n$ nucleons, $N_0$ being the
spherical magic number and $n$ a small even integer.
Classical PO families residing in the
shaded area in the octupole cavity give the shell effect similar to
spherical magic nuclei with particle number $N_0$.
With increasing $n$ (as illustrated in the upper panel with  $n'>n$),
larger octupole deformation is expected to keep the radius and the
curvature radius of the equatorial plane identical to the radius of
the spherical magic nucleus.}
\end{figure}

Figure~\ref{fig:pict2} illustrates the above argument from another
perspective.  If $n$ particles are
added to the spherical magic nucleus with particle number $N_0$,
system will favor the octupole shape with local symmetry
whose radius of the
equatorial plane coincides with the radius of spherical magic nucleus.
With more particles ($n'>n$) attached, larger octupole deformation
will occur.  The middle panels of Fig.~\ref{fig:pict2} display the
case of the stretched octupole parametrization,
for which one finds only local families of the quasi-periodic orbits
around the shaded area
in vicinity of the equatorial plane.  For the sphere + paraboloid
surface, POs form three-parameter families over the larger area
in the sphere part, as depicted by the shaded area in the right
panels of Fig.~\ref{fig:pict2}.  It would be also interesting to
examine how these difference in PO families might affect the octupole
shell effect.

\subsection{Cavity model with the stretched octupole parametrization}
\label{sec:cavity_oct}

\begin{figure}
\centering
\includegraphics[width=\linewidth]{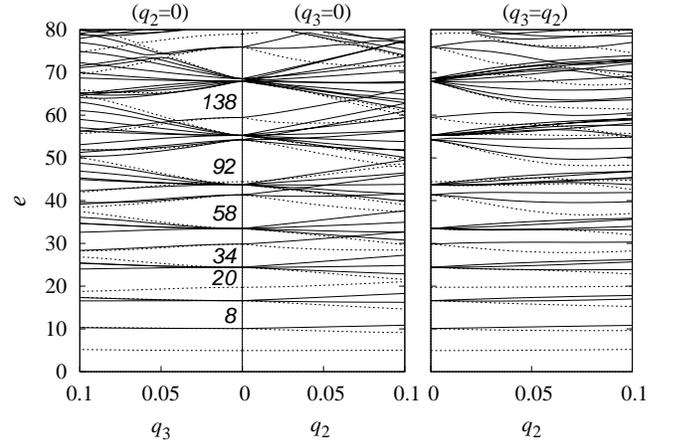} 
\caption{\label{fig:ndiag1}
Single-particle level diagram for spheroidal-octupole cavity model
with the stretched octupole parametrization.  Solid and broken curves
represent degenerate (magnetic quantum number $K\ne 0$) and
nondegenerate ($K=0$) levels, respectively.  The left panel shows
single-particle level diagrams for pure quadrupole and for pure octupole
deformations.  Spherical-shape magic numbers are indicated in italics.
In the right panel, the level diagram is shown for
deformations along $q_3=q_2$, close to the local symmetry condition.}
\end{figure}

Let us first consider the cavity model with the surface shape
parametrized by Eq.~(\ref{eq:shape_oct}).
Figure~\ref{fig:ndiag1} displays the single-particle spectra.
In the left panel, single-particle level diagrams for pure quadrupole
and pure octupole deformations are shown.
The integers put in the middle of the figure indicate the spherical
magic numbers, namely, the number of
levels below the energy gap.  They are close to the real nuclear
magic numbers, although slightly deviate from them mainly
due to the absence of spin-orbit coupling.  The doubly magic nuclei
\nuc{Sn}{132} and \nuc{Pb}{208} correspond to $(Z,N)=(58,92)$ and
$(92,138)$, respectively, in the cavity model.  The
degeneracies of levels at the spherical shape are broken with increasing
deformation without forming noticeable structure for both $q_2$
and $q_3$.

In the right panel of Fig.~\ref{fig:ndiag1}, level diagram is shown
along the deformations with $q_3=q_2$.  A remarkable feature for such
shapes is the existence of equally-spaced strongly bunched upward-right
levels, and it is expected that a significant amount of the spherical
shell effect will survive up to large deformations.  The spherical
symmetry is locally restored around the equatorial plane
as illustrated in the middle panels of
Fig.~\ref{fig:pict2} for the shape with $q_3\approx q_2$,
and the above shell
effect should be associated with the contribution of degenerate
families of POs in the semiclassical trace formulas
(\ref{eq:trace_g}) and (\ref{eq:trace_sce}).

\begin{figure}
\centering
\includegraphics[width=\linewidth]{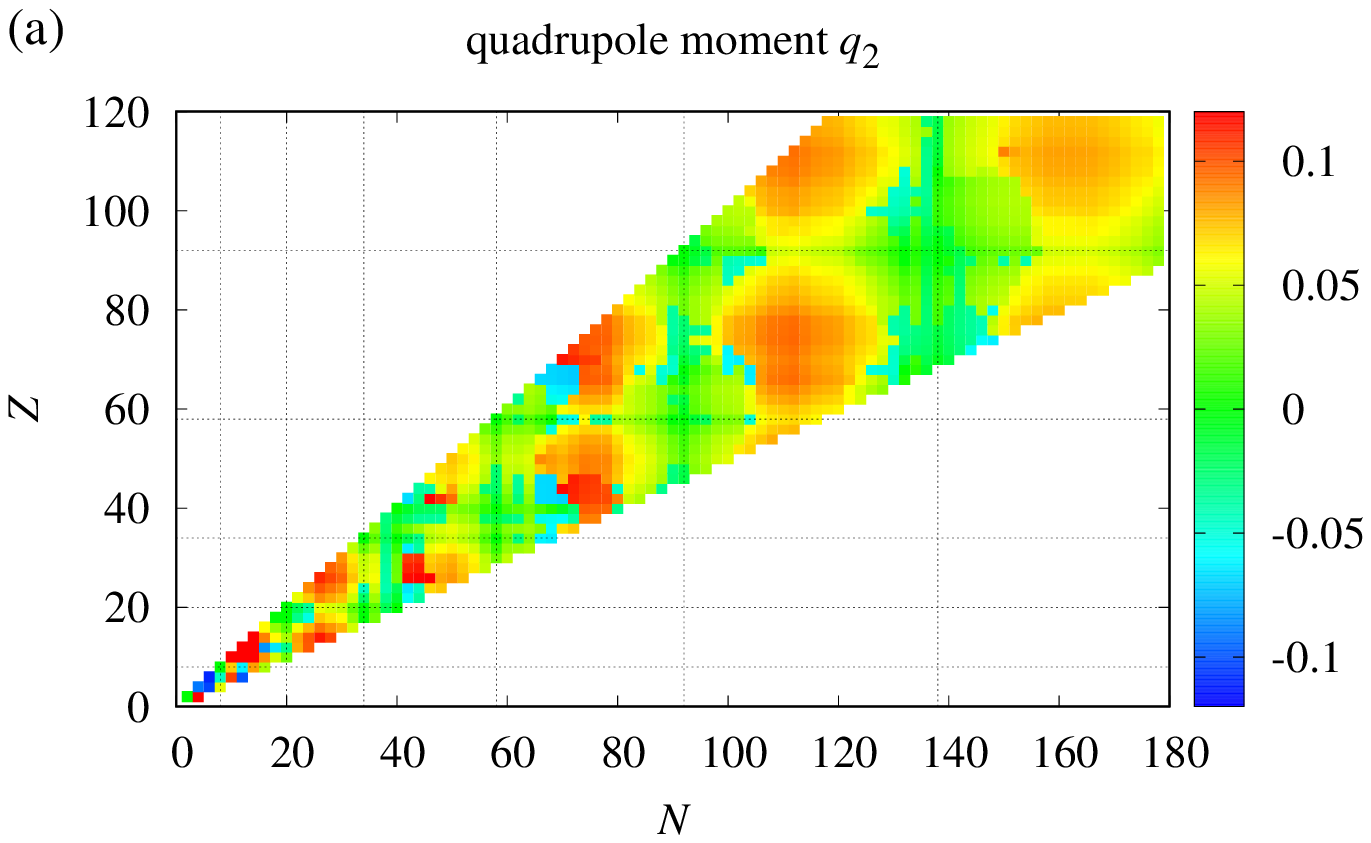} 
\\
\includegraphics[width=\linewidth]{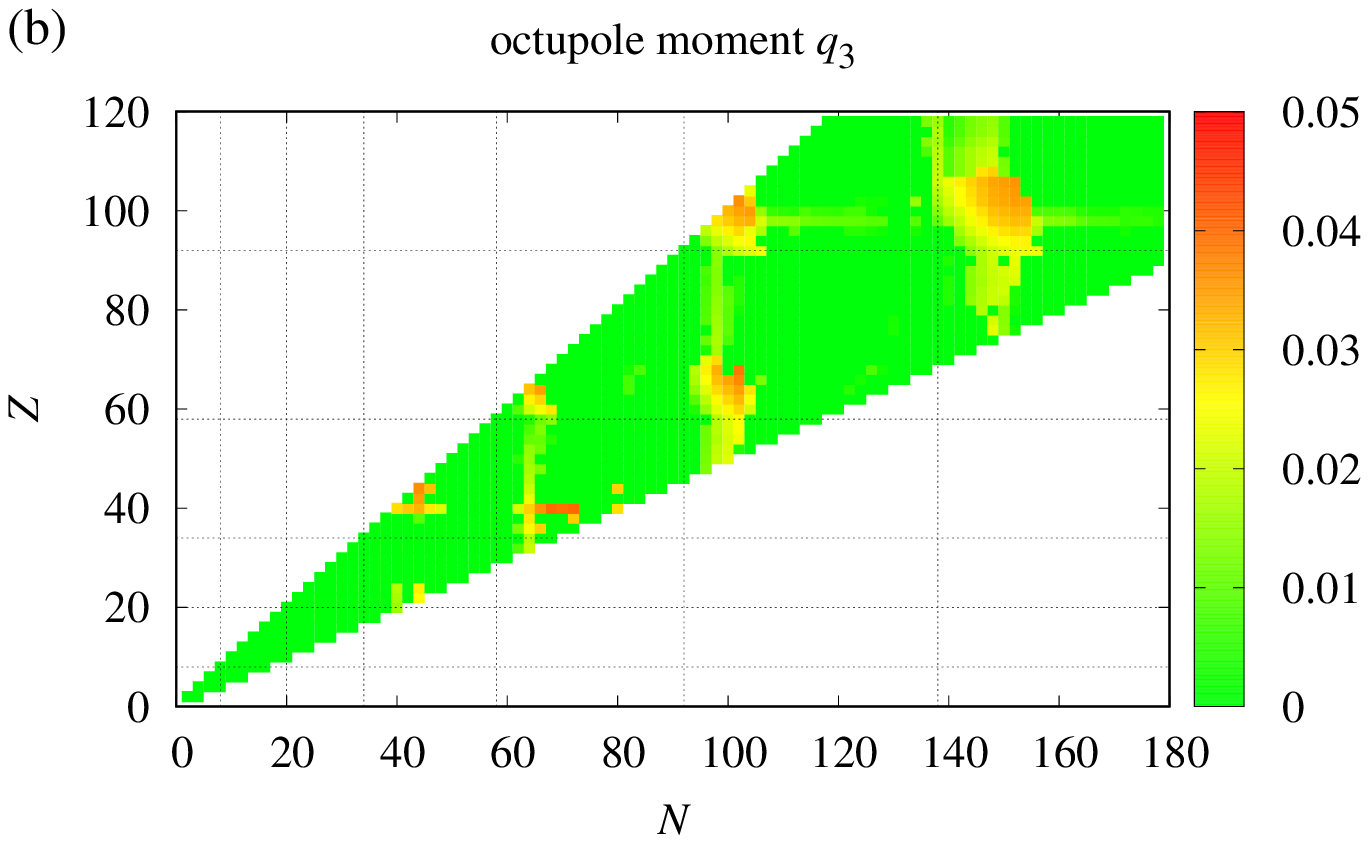} 
\caption{\label{fig:chart1_qm}
Ground-state quadrupole moment $q_2$ (upper panel) and octupole moment
$q_3$ (lower panel) in cavity potential model with stretched octupole
parametrization are shown on the $(N,Z)$ plane.  Horizontal and
vertical dotted lines indicate magic numbers, 8, 20, 34, 58,
92, 138, of the spherical cavity model.}
\end{figure}

With this single-particle spectra, energies of nuclei
given by Eq.~(\ref{eq:energy_cavity}) is calculated over the nuclear
chart (in the range $Z\leq N\leq 2Z$ which is approximately
corresponding to the region between proton and neutron drip lines) as
functions of $q_2$ and $q_3$, and ground-state deformations are
obtained.  The upper panel of Fig.~\ref{fig:chart1_qm} shows the
quadrupole moment $q_2$ of the ground-state shapes.  The horizontal and vertical
dotted lines indicate the spherical magic numbers.  One obtains
nearly spherical shapes along those magic lines.  In most of the other
regions, $q_2$ takes positive values, which reproduces the feature of
prolate-shape predominance in the ground-state
deformations \cite{BMBook2,Tajima02,Takahara12,Frisk90,Arita16}.  In
the lower panel of Fig.~\ref{fig:chart1_qm}, octupole moment $q_3$ is
plotted.  One will find that the octupole deformations systematically
appear at the \textit{north-east} neighbors of the
doubly closed-shell
configurations.  This qualitatively reproduces the results of
experiments and realistic calculations.  The essential mechanism to
explain the systematics of the octupole deformation seems to be
already involved in this simplified cavity model.

\begin{figure}
\centering
\includegraphics[width=\linewidth]{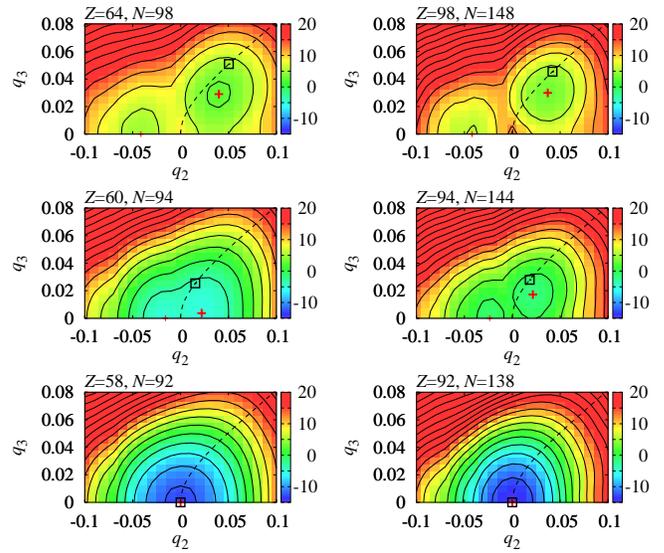} 
\caption{\label{fig:econt1}
Potential-energy surfaces for the
doubly magic nuclei $(Z,N)=(58,92),
(92,138)$ and their \textit{north-east} neighbors on the nuclear chart.
Cavity potential with the stretched octupole parametrization is
employed.  Contour curves of the shell-deformation energy
$E_{\textrm{sh-def}}$ in $(q_2,q_3)$ plane are drawn at intervals of
2~MeV.  The thick broken line represents the bifurcation deformation
for equatorial orbits.  Large cross represents the ground-state
deformation and small crosses are other local minima.  Square represents the
semiclassical guess of the ground-state deformation.}
\end{figure}

Figure~\ref{fig:econt1} shows the potential-energy surfaces of
doubly magic nuclei $(Z,N)=(58,92)$ and $(92,138)$ and their
\textit{north-east} neighbors on the nuclear chart.  Contour plot of the
shell-deformation energy (\ref{eq:energy_sh_def}) are shown as
functions of $(q_2,q_3)$, and the
ground-state deformations and other local minima are marked with the
cross symbols.  Adding a few neutrons and protons to the spherical
doubly magic nucleus, the system tends to take octupole shapes
accompanied by prolate quadrupole deformation.
The thick broken line indicate the bifurcation line of the equatorial
orbit where one has the local symmetry restoration.  As expected from the
POT, nuclear shape evolves along this bifurcation line
with increasing proton and neutron numbers.

In each panel of Fig.~\ref{fig:econt1}, the semiclassical guess of the
ground-state deformation is indicated by the square symbol, where the
radius of the equatorial plane coincides with the radius of the
spherical doubly magic nucleus.  The agreement with
the quantum result is not bad
but overestimating the deformation a little.  It might be improved by
taking the change of the Maslov index properly.

\subsection{Cavity model with spheroid-paraboloid parametrization}
\label{sec:cavity_oct2}

Next I examine the cavity model with spheroid-paraboloid
parametrization (\ref{eq:shape_wqs}).  Since the spherical symmetry is
maximally restored in the quadrupole-octupole deformed cavity, more
significant effect of the symmetry restoration is expected.

\begin{figure}
\centering
\includegraphics[width=\linewidth]{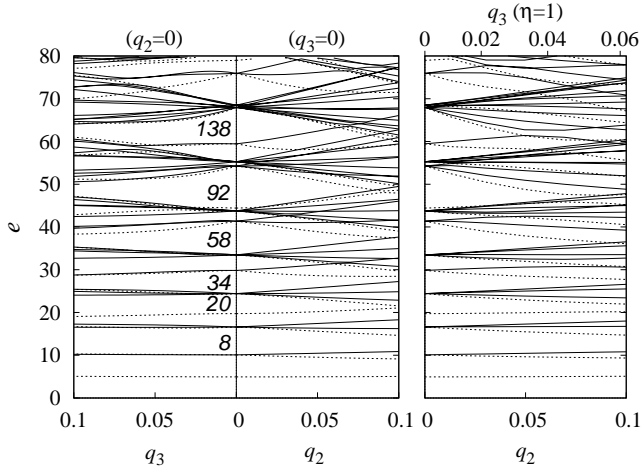} 
\caption{\label{fig:ndiag2}
Single-particle level diagram for the cavity potential model
with the spheroid-paraboloid parametrization.
The left panel is plotted in the same way as Fig.~\ref{fig:ndiag1}.
In the right panel, levels are plotted against $q_2$, where $q_3$
is varied with $q_2$ so that $\eta=1$ is satisfied.}
\end{figure}

Let us first look at the single-particle spectra in Fig.~\ref{fig:ndiag2}.
In the left panel, single-particle energies for pure quadrupole
and pure octupole shapes are shown.  There seems no noticeable
differences compared with those for the previous shape parametrization.
(The pure quadrupole shape is spheroidal and the diagram is equivalent
to the previous one.)  In the right panel, levels are plotted against
$q_2$, with $q_3$ varied with $q_2$ so that the spheroidal part of the
surface is kept spherical ($\eta=1$).  Again one finds bunches of
strongly degenerate upward-right levels, indicating the effect of
local spherical symmetry.

\begin{figure}
\centering
\includegraphics[width=\linewidth]{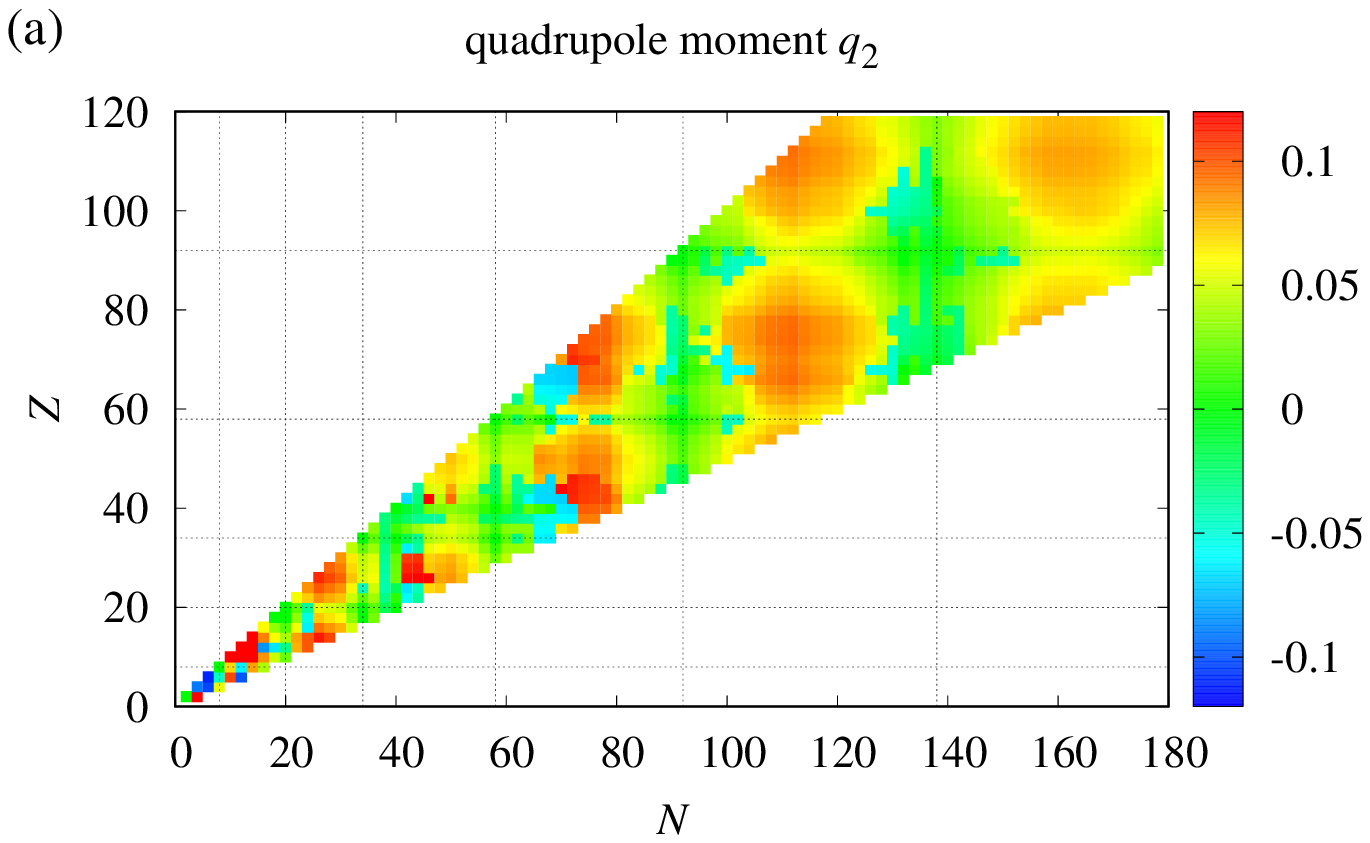} 
\\
\includegraphics[width=\linewidth]{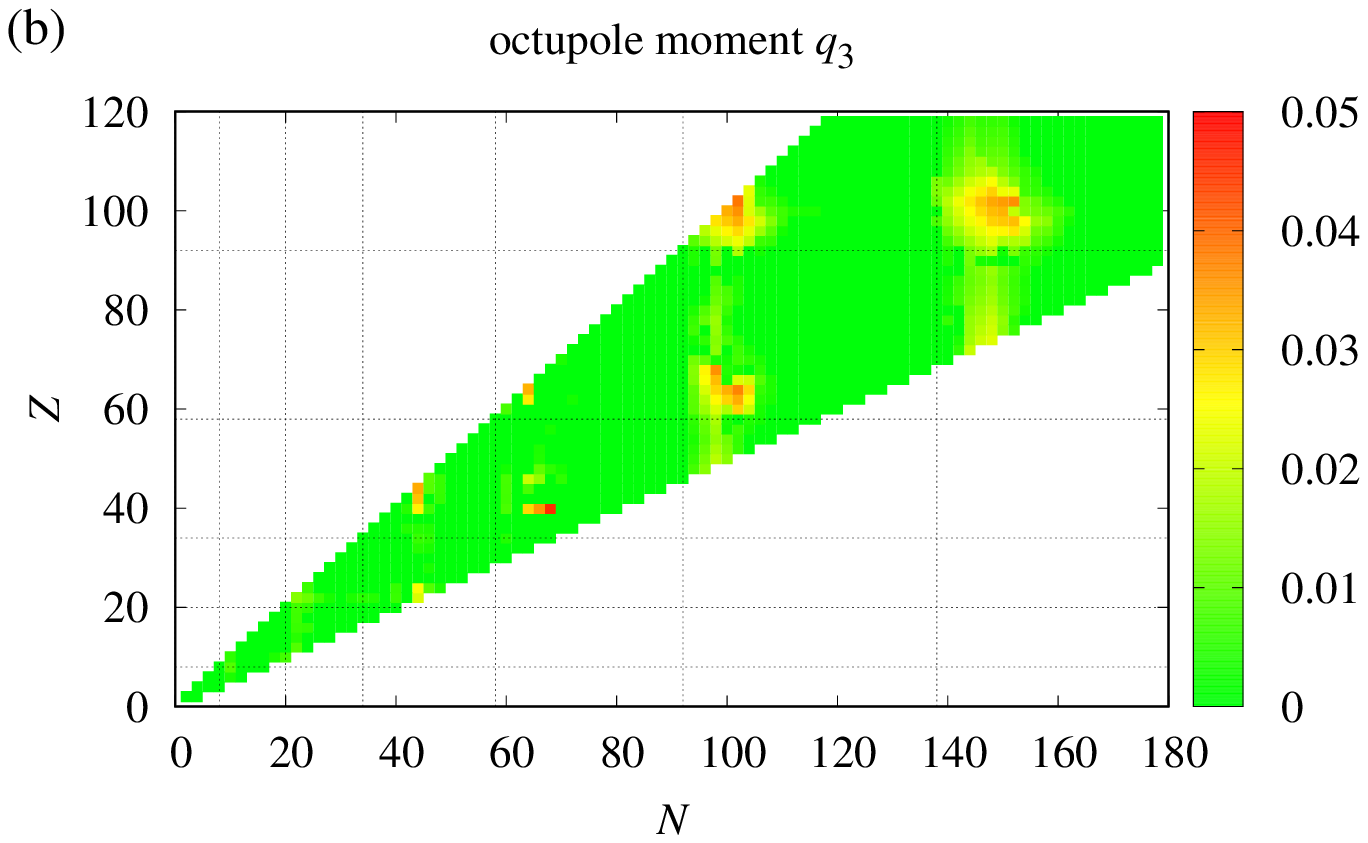} 
\caption{\label{fig:echart}
Same as Fig.~\ref{fig:chart1_qm} but with spheroid-paraboloid
parametrization.}
\end{figure}

The obtained ground-state deformations are shown in
Fig.~\ref{fig:echart}.  The results are qualitatively the same as the
previous parametrization.  The octupole deformations are found
systematically at the \textit{north-east} neighbors of each
doubly magic nucleus.

\begin{figure}
\centering
\includegraphics[width=\linewidth]{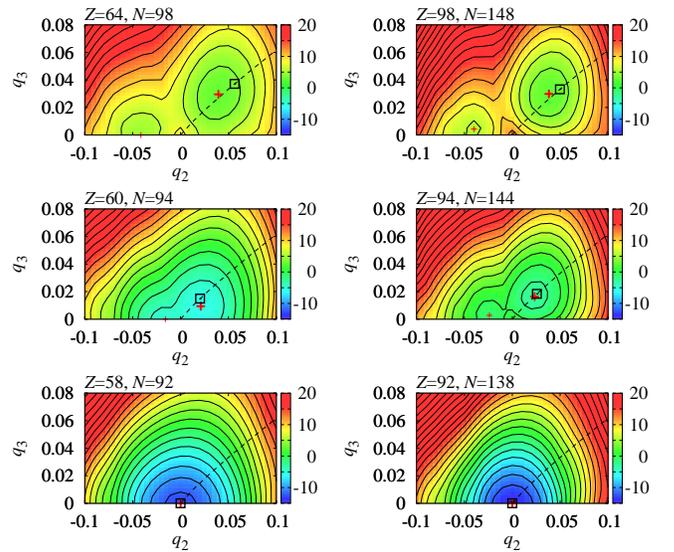} 
\caption{\label{fig:econt2}
Same as Fig.~\ref{fig:econt1} but with the spheroid-paraboloid
parametrization.  The thick broken line represents the shape with the
spheroidal parameter $\eta=1$.}
\end{figure}

Figure~\ref{fig:econt2} shows the potential-energy surfaces for the
same nuclei as those shown in Fig.~\ref{fig:econt1}.  In the present
parametrization, one has highly degenerate POs in the
spheroid part of the potential, and more significant shell effect due
to those orbits is expected.  The thick broken line represents the
shape where the spheroidal part of the surface is spherical and
involves triply degenerate POs.  In the same reason as
discussed in the previous section, one has shell energy minima when
the radius of the spheroid part is equal to that of the spherical
magic nucleus.  This condition gives the semiclassical guess of the
ground-state deformation, whose position is marked with the square
symbol in each panel of Fig.~\ref{fig:econt2}.  Since there is
no change of the Maslov index in this case, the agreement of the
semiclassical guess (\ref{eq:cond_emin}) with the quantum results
is almost perfect.

\begin{figure}
\includegraphics[width=.9\linewidth]{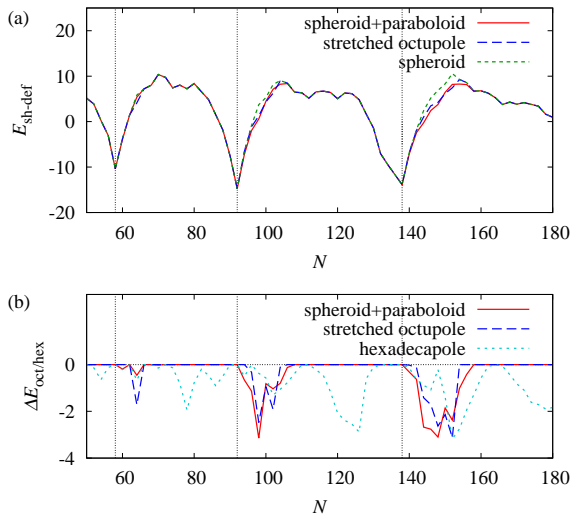} 
\caption{\label{fig:edef}
The upper panel (a) shows the shell-deformation energy of the ground
states of nuclei along the
$N=Z$ line.  Solid and broken curves represent the results for the
stretched octupole parametrization and spheroid+paraboloid
parametrization, respectively.  Dotted curve is for the pure
spheroidal deformation.
The lower panel (b) compares the octupole energy gain for the two
shape parametrizations.  The result for hexadecapole energy gain
is also shown.}
\end{figure}

Figure~\ref{fig:edef} compares the octupole energy gain
in the two parametrizations.
The upper panel displays the shell-deformation energy for neutron
or proton.  The lower panel displays the octupole
energy gain with respect to the
lowest energy in case of the spheroidal deformation alone.
One finds a systematic energy gain due to the octupole deformation
at just above the spherical shell closures for both parametrizations.

In terms of the order of semiclassical expansion, the PO
families with higher degeneracies in the spheroid+paraboloid
parametrization should provide more significant shell effect
than those in the stretched-octupole shape.
From the quantum-mechanical results shown in Fig.~\ref{fig:edef}(b),
one finds that the shell effect for the spheroid+paraboloid
parametrization becomes more significant with increasing $N$
as the semiclassical expansion becomes better.
The number of particle $N$ is limited to $N^{1/3}\lesssim 5$ for
nuclear systems and the difference is not so clear, but it will
become more pronounced in systems with much larger numbers of
particles, e.g., in metallic clusters.

Besides the octupole deformation, reflection-symmetric hexadecapole
deformation combined with quadrupole deformation can also cause
the same local symmetry.  I have also examined the shell
effect for the cavity potential models
by parametrizing the
quadrupole+hexadecapole shapes in the ways analogous to
Eq.~(\ref{eq:shape_oct}) or (\ref{eq:shape_wqs}).
The quantum-mechanical results show similar shell effects
just above the spherical closed-shell configurations as expected.
However, the effects were considerably smaller than those obtained for
the octupole shapes.  For a stretched hexadecapole shape, where $P_3$
in Eq.~(\ref{eq:shape_oct}) is replaced by $P_4$, the energy gain due
to the hexadecapole deformation is also plotted in
Fig.~\ref{fig:edef}(b).  To understand this difference, one will
have to consider the higher-order expansion of the action integral
around the PO and evaluate the so-called diffraction catastrophe
integral \cite{BkCatastrophe} as in the uniform
approximation \cite{Schom97}, which shall be left for the future subject.
I have also considered a shape analogous to
Fig.~\ref{fig:wqs} but another paraboloid is joined on the left
side in a symmetric way.  The obtained shell effect was smaller
than the case of a single paraboloid, possibly because of the
smaller parameter space occupied by a triangular orbit family in the
sphere part due to the truncations of the sphere on both sides.

\subsection{Oscillator-type potential model}
\label{sec:osc}

The cavity potential model is very useful when I make a semiclassical
analysis because of the simple form of the trace formula
(\ref{eq:trace_cavity}).  Without losing this simplicity, the
potential can be made more realistic by generalizing it to the radial
power-law potential \cite{Arita12,AriMuk14}.
The deformed power-law potential with the shape (\ref{eq:shape_oct}) is
expressed as
\begin{equation}
V(\br)=U_0\left(\frac{r}{r_s(\theta)}\right)^\alpha.
\end{equation}
The power parameter $\alpha$ controls the radial dependence of the
potential.  The limit $\alpha\to\infty$ corresponds to the cavity and
$\alpha=2$ corresponds to the harmonic oscillator.  It is well known that
all degenerate levels in the spherical harmonic-oscillator potential
consist of identical parities, respectively.
In the following, let us consider
the case $\alpha=2$.  This will help us verify the importance of
the $\varDelta l=3$ mixing for octupole deformation.

\begin{figure}
\centering
\includegraphics[width=\linewidth]{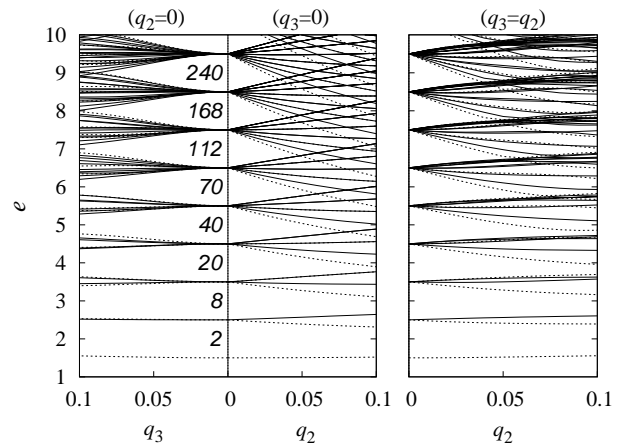} 
\caption{\label{fig:ndiag_a02}
Single-particle level diagrams same as Fig.~\ref{fig:ndiag1}
but for the radial power-law potential model with $\alpha=2$.
The numbers in italics put in the plots indicate the magic numbers of
the spherical harmonic-oscillator potential model.}
\end{figure}
Single-particle level diagrams are shown in Fig.~\ref{fig:ndiag_a02}.
Because of no octupole matrix elements between levels within each shell,
the spectrum is stiffer against pure octupole deformation
compared with the cavity case.

\begin{figure}
\centering
\includegraphics[width=\linewidth]{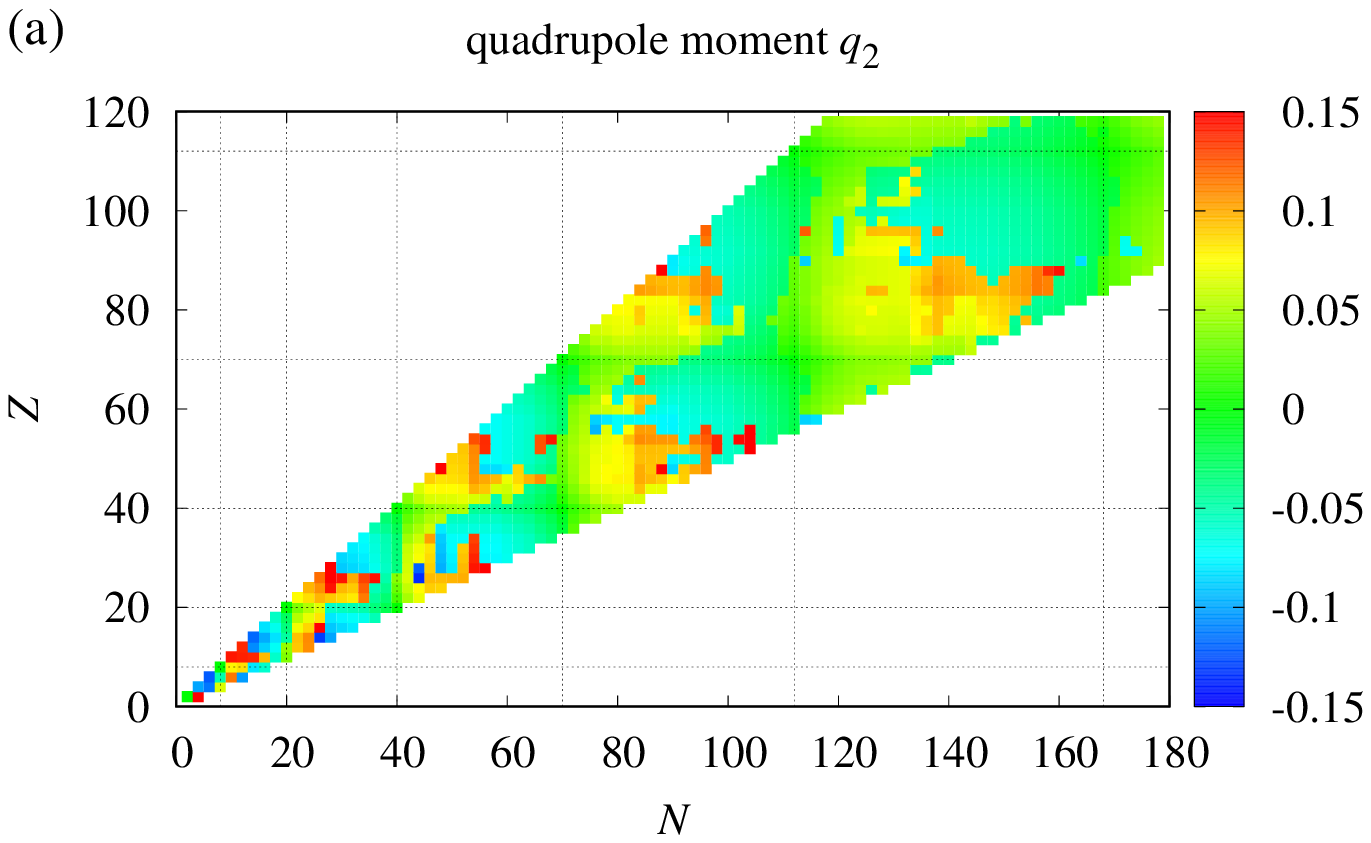} 
\\
\includegraphics[width=\linewidth]{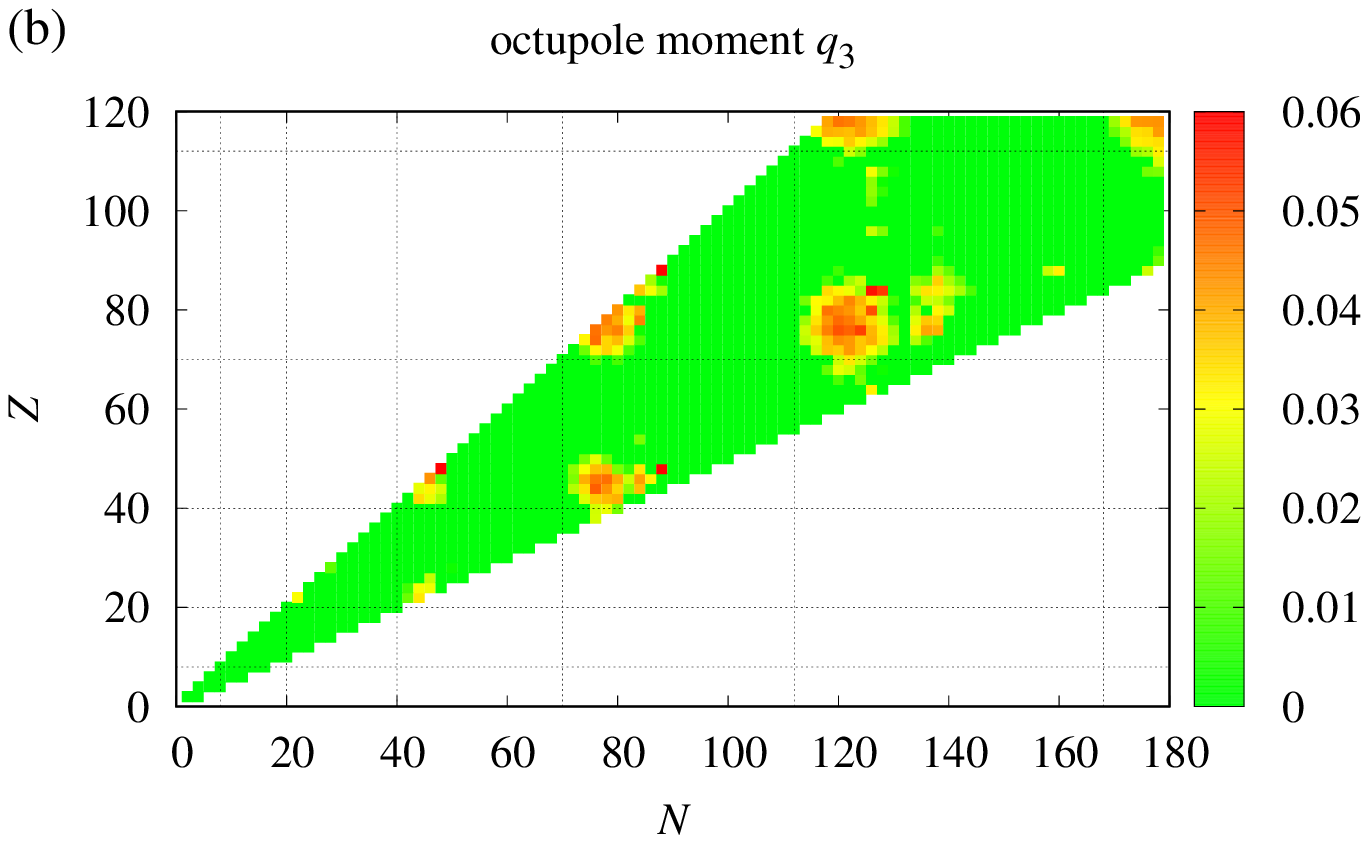} 
\caption{\label{fig:chart02_qm}
Same as Figs.~\ref{fig:chart1_qm} and \ref{fig:echart} but for the
radial power-law potential
model with the power parameter $\alpha=2$.  Spherical magic numbers
are 8, 20, 40, 70, 112, 168, $\cdots$.}
\end{figure}

Ground-state deformations are determined by minimizing the energy
(\ref{eq:energy_rpl}) as before.  In the upper panel of
Fig.~\ref{fig:chart02_qm}, ground-state quadrupole moment is shown.
In this case, prolate and oblate shapes appear approximately at
equal rates.  Looking at the octupole moment in the lower panel
of Fig.~\ref{fig:chart02_qm}, one again finds systematic appearance of
octupole deformations just above the doubly magic nuclei.
This is related to the shell effect associated with the PO
bifurcation which occur for certain combinations of quadrupole and
octupole deformations \cite{Arita95}.

\begin{figure}
\centering
\includegraphics[width=.6\linewidth]{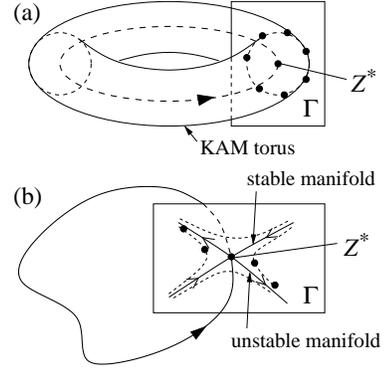} 
\caption{\label{fig:pss}
Illustration of Poincar\'{e} surface of section plot around
(a) stable and (b) unstable POs.
$Z^*$ represents the section of the PO.}
\end{figure}

In investigating the classical PO bifurcation, Poincar\'{e} surface of
section (PSS) plot (see Sec.~\ref{sec:pot}) around the PO
is useful.  Stable (regular) trajectories are confined on the
so-called KAM torus and the PSS plots for such trajectory
accumulate on a
closed curve corresponding to the intersection of the torus and the
surface of section $\Gamma$.  Thusm, concentric structures are formed
in the PSS plot around the stable PO, as illustrated in
Fig.~\ref{fig:pss}(a).  On the other
hand, the PSS for an unstable (chaotic) trajectory fills a certain
region of the surface $\Gamma$ in a random manner.  Unstable
POs are generally buried in chaotic region, but 
just after their birth through the bifurcations, they can be easily
found as the intersection of the stable and unstable manifolds,
as illustrated in Fig.~\ref{fig:pss}(b).

\begin{figure}
\centering
\includegraphics[width=\linewidth]{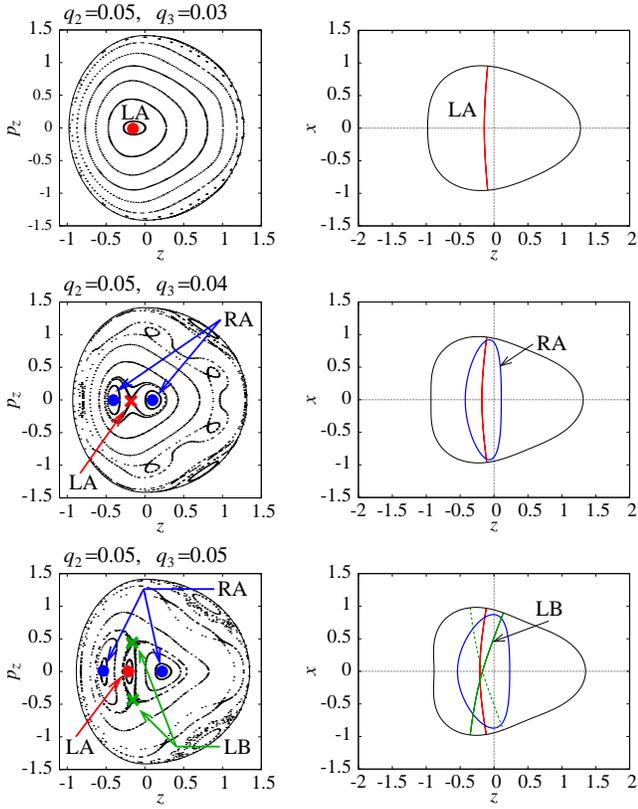} 
\caption{\label{fig:pmap02}
Poincar\'{e} surface of section (PSS) plots (left panels) and
simple classical POs (right panels) for the radial
power-law potential model with $\alpha=2$, $q_2=0.05$ and several
values of $q_3$.
The PSS plots are for the planer classical trajectories in $(x,z)$
plane with the section $\Gamma: x=0$.  Stable and unstable fixed
points corresponding to the stable and unstable POs are marked with
solid circles and crosses, respectively.
As for the names of the POs, L and R stand for the acronyms
``linear'' (or ``librating'') and ``rotating'', respectively.}
\end{figure}

Figure~\ref{fig:pmap02} shows the PSS plots
and the relevant classical POs.  
In the upper panels, at $q_3=0.03$, one has a stable linear orbit LA which
forms a one-parameter family with respect to the rotation about the
symmetry axis.  With increasing $q_3$, one finds in the middle panels,
at $q_3=0.04$, a new 1-parametric orbit RA which has emerged through the
bifurcation of LA, after which the orbit LA becomes unstable.
In the bottom panels, at $q_3=0.05$, another new
linear 1-parametric orbit LC has emerged through the second
bifurcation of LA.
\begin{figure}
\centering
\includegraphics[width=.75\linewidth]{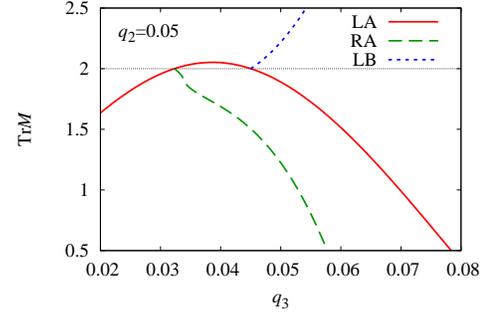} 
\caption{\label{fig:trm}
Traces of the symmetry-reduced monodromy matrices for the orbit
LA and its bifurcation daughters, plotted as the functions of
the octupole parameter $q_3$ with fixed quadrupole parameter $q_2=0.05$.}
\end{figure}
For these orbits, symmetry-reduced monodromy matrices are
$(2\times 2)$ real symplectic ones, and their eigenvalues appear
either in a pair $(e^{i\alpha},e^{-i\alpha})$ for stable orbits or
$(e^{\beta},e^{-\beta}),~ (-e^{\beta},-e^{-\beta})$ for unstable
orbits, where $\alpha$ and
$\beta$ are real numbers.  Using these properties,
the stability factor (\ref{eq:stability}) is expressed as
\begin{equation}
\frac{1}{\sqrt{|\det(I-\tilde{M}_{\rm PO})|}}
=\frac{1}{\sqrt{|2-\Tr\tilde{M}_{\rm PO}|}}.
\end{equation}
$|\Tr\tilde{M}_{\rm PO}|=|2\cos\alpha|<2$ for a stable PO and
$|\Tr\tilde{M}_{\rm PO}|=2\cosh\beta>2$ for an unstable PO, and the
bifurcation of PO occurs at $\Tr\tilde{M}_{\rm PO}=2$ where the
eigenvalues of $\tilde{M}_{\rm PO}$ become $1$ ($\alpha=0$ or
$\beta=0$).
Thus, the history of the bifurcations can be clearly examined by
looking at the trace of the monodromy matrix.
Figure~\ref{fig:trm} shows the trace of the
symmetry-reduced monodromy matrix as the function of the octupole
parameter $q_3$, with quadrupole parameter is fixed to $q_2=0.05$.
With increasing $q_3$, the orbit LA causes bifurcation
and a new orbit RA emerges at $q_3=0.0323$.  Then, LA causes the
second bifurcation and another new orbit LB emerges at $q_3=0.0451$.
The occurrence
of such successive bifurcations in close proximity is known as the
codimension-2 bifurcation \cite{Schom98,AriBra08A}.  It indicates a
restoration of dynamical symmetry with higher dimension, and one
can expect more significant influence on the shell effect than the simple
bifurcations.

In the right panel of Fig.~\ref{fig:ndiag_a02}, single-particle diagram
for the deformation $q_3=q_2$ is shown, which approximately along the
bifurcation points.  One finds bunched upward levels
in the same manner as the cavity models,
and the considerable amount of the spherical shell effect
is expected to survive for finite octupole deformation along
$q_3\approx q_2$.

\begin{figure}
\centering
\includegraphics[width=\linewidth]{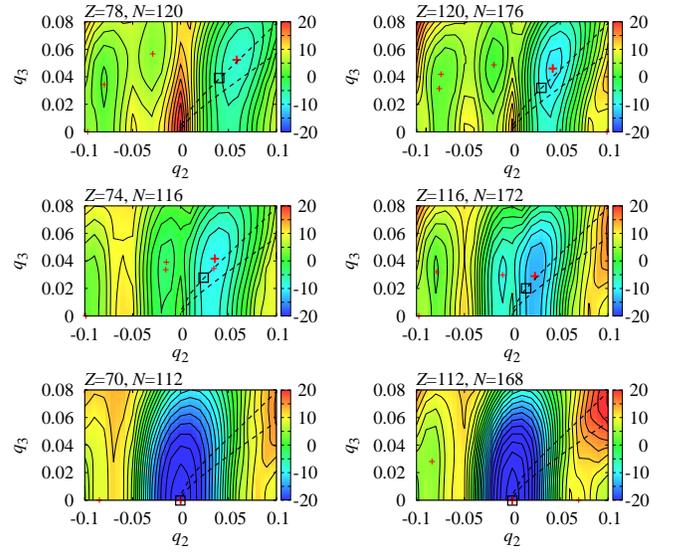} 
\caption{\label{fig:econt_a02}
Same as Fig.~\ref{fig:econt1} but for the radial power-law
potential model with the power parameter $\alpha=2$.
Results for doubly magic nuclei $(N,Z)=(78,112)$, (112,168) and some
of their upper-right neighbors in the nuclear chart are shown.
Two thick broken curves in each panel represent the first and the
second bifurcations of the symmetric self-retracing orbit LA.
}
\end{figure}

Figure~\ref{fig:econt_a02} shows the potential-energy surfaces for
doubly magic configurations $(Z,N)=(70,112)$ and $(112,168)$, and
their upper-right neighbors on the nuclear chart.  Two broken curves
in each panel indicate the lines of two bifurcation points of the
orbit LA which generate the orbits RA and LB, respectively.  One sees
that the ground-state deformation is approximately evolving along
these bifurcation lines as the particle numbers deviate from the
spherical magic numbers.  The semiclassical prediction of the optimum
shape can be made in the same way as for the cavity model, by generalizing
the wave number $k$ and the orbit length $L_{\rm PO}$ into the scaled
energy $\mathcal{E}=(E/U_0)^{1/\alpha+1/2}$ and the scaled action
$\tau_{\rm PO}=S_{\rm PO}/\hslash\mathcal{E}$,
respectively \cite{Arita12}.  In this case, the semiclassical guess
of the optimum shape underestimate the deformation a little.  This
might be also related to the change of the Maslov index but in the way
different from the case of the cavity model.

\begin{figure}
\includegraphics[width=.8\linewidth]{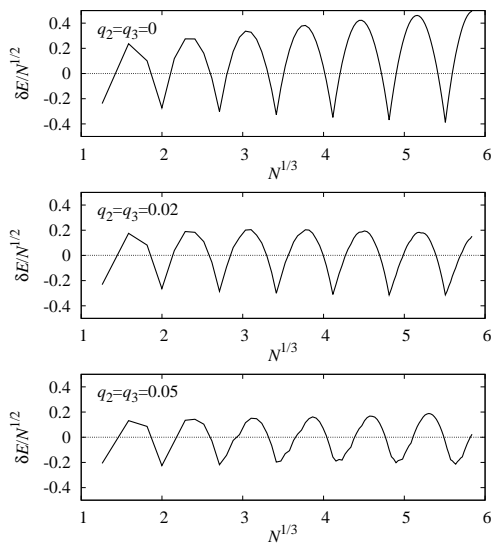} 
\caption{\label{fig:sce_a02}
Shell energies for the radial power-law potential model along the
deformations $q_2=q_3$, which are
approximately corresponding to bifurcation deformations.}
\end{figure}

Figure~\ref{fig:sce_a02} shows the shell
energy as function of particle number.  Along the bifurcation line,
regular oscillating structure similar to that for the spherical shape
is preserved up to large deformations.  This can be clearly understood
by the significant contribution of the bifurcating PO, relevant to the
gross shell structure.

\section{Summary}
\label{sec:summary}

Ground-state octupole deformations are systematically investigated by
the simple cavity models, taking into account the
quadrupole and octupole
shape degrees of freedom in two different ways of parametrizations.
The systematic appearance of octupole deformations just above the
spherical closed-shell configurations are understood as the gross
shell effect related to the classical PO contributions enhanced by the
local symmetry restorations.

The above systematics can be also reproduced without the help of
$\Delta l=3$ mixing in the oscillator-type
potential model.  This strongly suggests the significance of the
gross shell effect as playing an essential role in the
mechanism causing octupole deformation.

In spite of the extreme simplification of the model, it helps our
qualitative understanding of the microscopic mechanism for the
breaking of reflection symmetry.  The mechanism described in this
paper should also apply to more realistic mean-field potential models.

For the breaking of reflection symmetry,
the importance of nonaxial octupole degrees of freedom is
also suggested \cite{HamMot91,Frisk94,Yang22A,Yang22B}.
The effect of the point-group symmetry and the gross shell effect in
tetrahedral deformation is of particular
interest \cite{Dudek02,Dudek18,AriMuk14}.
The extension of this work to other exotic shape degrees of freedom
would be also an interesting subject for the future study.

As discussed in the end of Sec.~\ref{sec:cavity_oct2},
reflection-symmetric hexadecapole
deformation combined with quadrupole deformation can also cause
the same local symmetry.
The quantum-mechanical results show similar shell effects
just above the spherical closed-shell configurations, but they are
considerably smaller than those obtained for the octupole shapes.
Possible semiclassical reasons have been given but they apply only to
the current schematic models, and more careful study on the competition with
hexadecapole shape degree of freedom might be necessary when one
consider the breaking of reflection symmetry in realistic
models.

\acknowledgments

The author would like to thank the members of the Nagoya Nuclear Physics
Colloquium for helpful discussions.
Part of the numerical calculations are carried out at the Yukawa
Institute Computer Facility.

\bibliographystyle{apsrev4-2}
\bibliography{refs}
\end{document}